# Development of the optical system for the SST-1M telescope of the Cherenkov Telescope Array observatory

K. Seweryn[h], W. Bilnik[k], J. Błocki[g], L. Bogacz[e], T. Bulik[d], F. Cadoux[a], A. Christov[a], M. Curyło[g], D. della Volpe[a], M. Dyrda[g], Y. Favre[a], A. Frankowski[c], Ł. Grudnik[g], M. Grudzińska[d], M. Heller[a], B. Idźkowski[b], M. Jamrozy[b], M. Janiak[c], J. Kasperek[k], K. Lalik[k], E. Lyard[f], E. Mach[g], D. Mandat[m], A. Marszałek[h,b], J. Michałowski[g], R. Moderski[c], T. Montaruli[a], A. Neronov[f], J. Niemiec[g], M. Ostrowski[b*], P. Paśko[h], M. Pech[m], A. Porcelli[a], E. Prandini[f], P. Rajda[k], M. Rameez[a], E. jr Schioppa[a], P. Schovanek[m], K. Skowron[g], V. Sliusar[i], M. Sowiński[g], Ł. Stawarz[b], M. Stodulska[b], M. Stodulski[g], S. Toscano[f,l], I. Troyano Pujadas[a], R. Walter[f], M. Więcek[k], A. Zagdański[b], K. Ziętara[b], P. Żychowski[g]
and T. Barciński[h], M. Karczewski[h], J. Nicolau–Kukliński[h], Ł. Płatos[h], M. Rataj[h], P. Wawer[h], R. Wawrzaszek[h] for the CTA Consortium[†].

[a] *DPNC – Université de Genève, Genève, Switzerland;* [b] *Astronomical Observatory, Jagiellonian University, Kraków, Poland;* [c] *Nicolaus Copernicus Astronomical Centre, Polish Academy of Sciences, Warsaw, Poland;* [d] *Astronomical Observatory, University of Warsaw, Warsaw, Poland;* [e] *Department of Information Technologies, Jagiellonian University, Kraków, Poland;* [f] *ISDC, Observatoire de Genève, Université de Genève, Versoix, Switzerland;* [g] *Instytut Fizyki Jądrowej im. H. Niewodniczańskiego Polskiej Akademii Nauk, Kraków, Poland;* [h] *Centrum Badań Kosmicznych Polskiej Akademii Nauk, Warsaw, Poland;* [i] *Astronomical Observatory, Taras Shevchenko National University of Kyiv, Kyiv, Ukraine;* [k] *AGH University of Science and Technology, Kraków, Poland;* [l] *Vrije Universiteit Brussels, Brussels, Belgium;* [m] *Institute of Physics of the Czech Academy of Sciences, Prague, Czech Republic*
E-Mail:kseweryn@cbk.waw.pl

The prototype of a Davies-Cotton small size telescope (SST-1M) has been designed and developed by a consortium of Polish and Swiss institutions and proposed for the Cherenkov Telescope Array (CTA) observatory. The main purpose of the optical system is to focus the Cherenkov light emitted by extensive air showers in the atmosphere onto the focal plane detectors. The main component of the system is a dish consisting of 18 hexagonal mirrors with a total effective collection area of 6.47 m$^2$ (including the shadowing and estimated mirror reflectivity). Such a solution was chosen taking into account the analysis of the Cherenkov light propagation and based on optical simulations. The proper curvature and stability of the dish is ensured by the mirror alignment system and the isostatic interface to the telescope structure. Here we present the design of the optical subsystem together with the performance measurements of its components.

---

[*]Speaker

[†] Full consortium author list at http://cta-observatory.org

*Development of an optical system for the SST-1M telescope of the Cherenkov Telescope Array observatory*

## 1. Introduction

Small-size telescopes (SSTs) are designed to cover the highest energy range of the Cherenkov Telescope Array (CTA) observatory [1]. A consortium of Polish and Swiss institutions is developing a telescope prototype that takes advantage of the standard and well proven Davies-Cotton optical design. It will be equipped with an innovative and fully digital camera based on detectors utilizing Silicon Photomultipliers (SiPM) technology.

The SST-1M has a focal length of 5.6 m, a pixel angular aperture equals to 0.24°, a geometrical field of view of 9°, and a reflective dish with a diameter of 4 m. The dish is composed of 18 hexagonal mirror facets with size of 78 cm flat-to-flat. The theoretical total effective collection area is 9.42 m$^2$ reduced by structure shadowing and reflectance of mirrors to 6.47 m$^2$.

In this paper the description of optical system is provided supplemented by recent hardware developments.

## 2. Design of the Optical system

The optical system of the SST telescope provides the optical path for the Cherenkov light emitted by showers in the atmosphere to be focused on the focal plane and recorded by the DigiCam camera [2, 3]. The technical representation of this path fits into SST-1M telescope structure [4] and by design fulfills the CTA scientific, technical and economical requirements [5]. The SST-1M optical system consists of: (i) 18 hexagonal mirror facets that form the telescope optical dish, (ii) the mirror alignment system that allows to control the curvature of the dish and (iii) the mirror alignment fixing set actuating each particular mirror in specified direction. All of these components are shown in figure 1 and discussed in this proceedings.

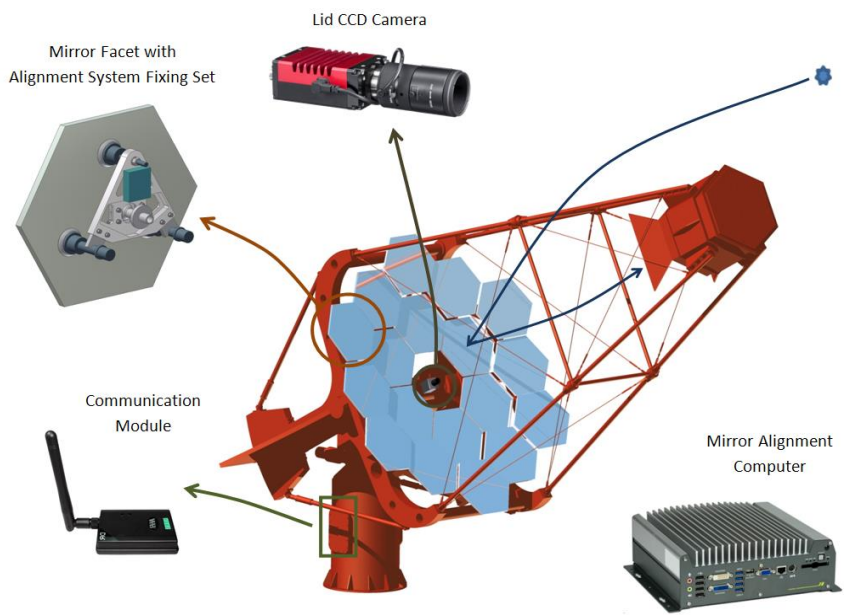

*Figure 1 Optical System of SST-1M*

*Development of an optical system for the SST-1M telescope of the Cherenkov Telescope Array observatory*

The 18 hexagonal facets are 780mm flat-to-flat, arranged in two concentric rings. They form a reflecting spherical dish with a diameter of about 4m (with a space of 2 cm between the tiles). The central facet is obscured by the camera and therefore in that place the Lid CCD camera is installed. The key parameters of the optical system are listed in Table 1.

Table 1. Key parameters of SST-1M optical system

| Focal Length | 5600±5mm | |
|---|---|---|
| f/D | 1.4 | |
| Dish Diameter | 4m | |
| Mirror Area | 9.42m² | Total collection area without shadowing |
| Mirror Effective Area | 6.47m² | Including shadowing and mirror reflectivity |
| Hexagonal Mirror Facets | 780±3mm flat to flat | |
| PSF (on-axis) | 0.07° | |
| PSF (@4deg off-axis) | 0.21° | The CTA requirement for PSF is 0.25° |

## 2.1 Optical point spread function

The point spread function (PSF) represents the optical quality of the telescope. The PSF is here defined to be the angular diameter of a circle containing 80% of the photons emitted by a point source at infinity and reflected by the dish, denoted by D80 (figure 2). Given the fact that the spot profile at the focal plane is not necessarily symmetrical, we define D80 by starting from the centre of gravity of the photon distribution and integrating the signal in larger and larger circles until 80% of the total emission is reached. For the camera field of view of 9°, the PSF needs to be determined for light rays coming from a source located at $\sqrt{0.8} \times 4.5°$ i.e. at 4°off-axis to demonstrate that the CTA requirements are fulfilled. (The factor in the square root comes from the 80% requirement on the FoV area in the above).

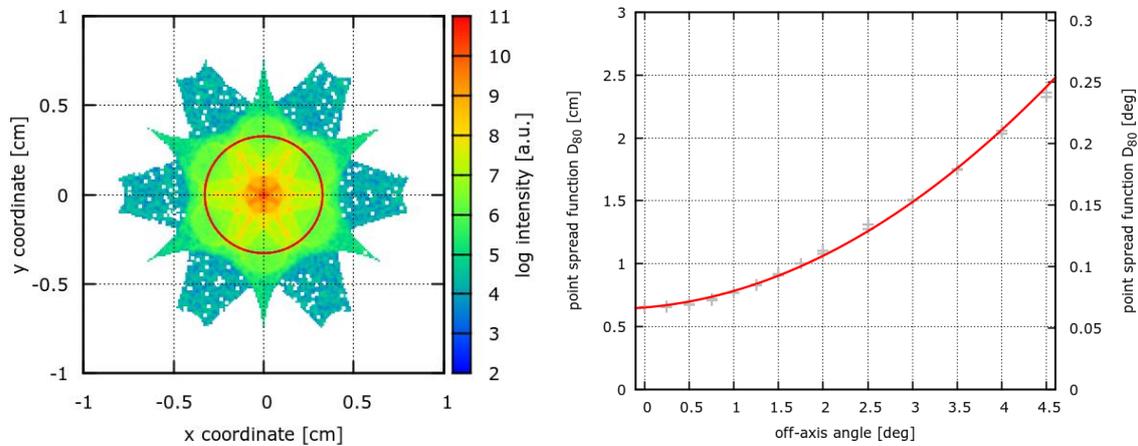

*Figure 2 Left: PSF shape for the whole dish - PERFECT case. Red circle corresponds to D80(f) = 0.654 cm. Right: PSF dependence on off-axis angle - PERFECT case. Gray marks show difference for horizontal and vertical rays.*

*Development of an optical system for the SST-1M telescope of the Cherenkov Telescope Array observatory*

## 2.2 Optical time spread

The telescope must focus light (over 80% of the required camera field of view) with an rms optical time spread of <1.5 ns. The results of the simulations presented in the last section have been used to determine the distribution of the photon arrival time at the focal plane. For 80% of the camera field of view of 9°, i.e. within 4° of the telescope axis, the largest time spread is 0.244 ns for on-axis rays.

## 3. Mirror facets

The mirror facet for SST-1M telescope was designed to be spherical with radius of curvature equal to 11.2 m. Each hexagonal mirror is 780 mm flat-to-flat. The reflectivity of the mirror is designed to be >0.87 in range of 300 nm-600 nm. Such solution gives the margin with respect to 85% reflectivity requirement. In operational conditions 80% of light reflected by each mirror facet must be contained in a circle with a diameter of 8.1mm (1/3 of 24.3mm pixel size) located in the focal plane. Mirrors also have to fulfill mechanical and thermal requirements. The baseline solution of mirror technology takes advantage of glass substrate and standard $AlSiO_2$ coating. This is a well-known solution, which may achieve 90% reflectivity. Nonetheless, the technology is very time-consuming, so the mass production schedule should be considered very carefully according to the manufacturer's capacity. In parallel, Sheet Moulding Compound (SMC) composite mirrors are under development at CBK PAN [6]. The substrate of the mirror is formed during a compression of SMC material in a mould on a press. SMC is a low-cost, lightweight, widespread and semi-fabricated product used for compression moulding. Both mirror facet designs are presented in figure 3.

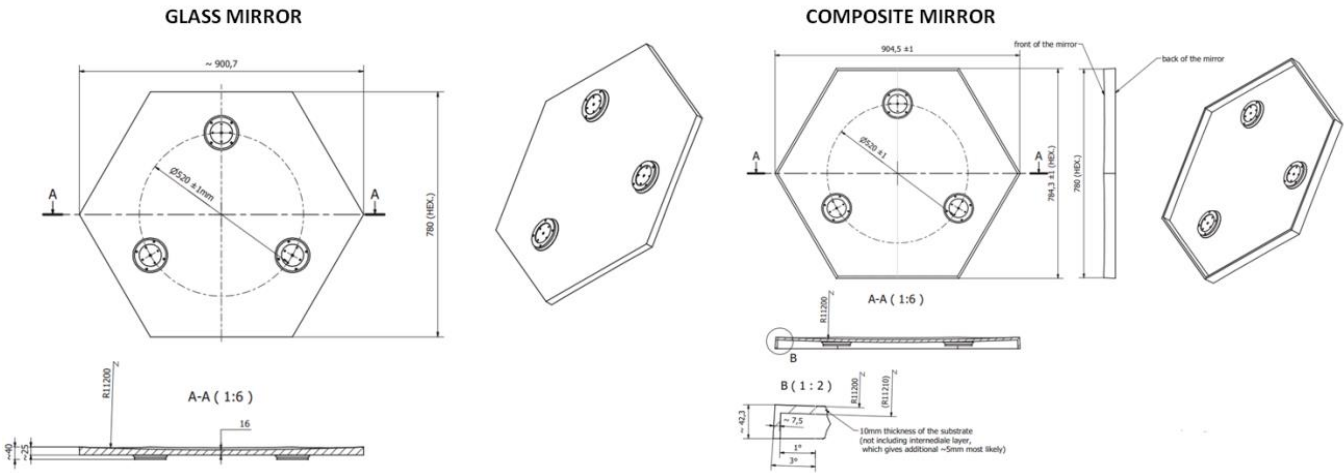

*Figure 3 The mechanical interface drawing for glass (left) and SMC composite (right) mirror facets for the SST-1M.*

## 4. Mirror Alignment

The automatic alignment process positions the mirror focus on the detector using optical feedback. The mirror focuses the light from a distant source at its focal plane, creating a spot on the screen. The coordinates of the position of the spot $p=[x,y]^T$ are related to the actuators' positions $q=[q_1,q_2]^T$ by one-to-one map $f:D_q \rightarrow D_p$, $q=f(p)$, where $D_q \rightarrow R^2$, $D_p \rightarrow R^2$ are the ranges of the

*Development of an optical system for the SST-1M telescope of the Cherenkov Telescope Array observatory*

actuator and spot positions, respectively. It can be demonstrated that the mapping is approximately linear in the operation range of the actuators, therefore the Jacobian matrix $J=\partial f(q)/\partial q$ is constant. The actuators' control system allows for operation in a sequence of displacements defined as differences of positions, i.e. $\Delta p=p_n-p_{n-1}$ Hence, the displacement of the spot is related to these displacements by the relation $\Delta p=J\Delta q_n$. Once the reference position of the spot $p_{ref}$ and the position tolerance $\varepsilon$ are defined, then the alignment control can be obtained as the sequence of desired actuator displacements $\Delta q_n=kJ^{-1}(p_{n-1}-p_{ref})$, iterated from position $p_0$ until the spot reaches the desired position $p_{ref}$ with the specified tolerance, i.e. $|p_i-p_{ref}|<\varepsilon$. The constant k is the control gain which controls the strength of the adjustment from gentle ($k\approx 0$) to aggressive ($k\approx 1$). In the case of k=0, the control target is achieved in one step. The main difficulty in adjusting the parameters in $\Delta q_n$ is the necessity of computation of the Jacobian matrix, which requires the precise measurement of the geometric properties of the system, for each mirror separately.

### 4.1 Mirror Alignment Fixing Set

The Mirror Alignment Fixing Set primary function is threefold: (i) mounting of the mirror in secure and reliable way into the telescope structure, (ii) not transferring the thermal stresses between telescope structure and mirrors and (iii) actuating the mirror facet orientation defined by Mirror Alignment System. In addition, the system allows for easy mounting and removal of the mirrors from telescope structure as well as manual pre-alignment of mirror (both mirror facet orientation and its radial distance from camera). The system consists of a triangle with ball joint, automatic actuators and control electronics (figure 4).

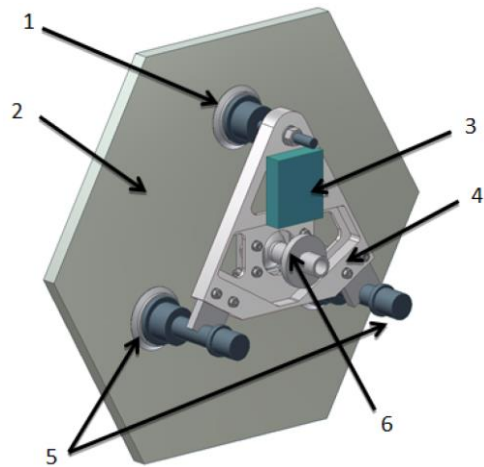

The mirror fixation is based on a 3-point isostatic mounting enabling secure fixing to the telescope structure without transferring thermal stresses from telescope structure. One interface is a so-called fixed point of the mirror, while two others are linear actuators (details in subsection 2.6). The orientation of the mirror facet is stable with arc minute level and controllable in two angular directions allowing to precisely and automatic positioning each mirror spot on desired point on focal plane. Each mirror is fixed to the interfaces via steel pads glued on its rear surface. The pads are uniformly distributed on a circle having its center in the middle of the mirror.

*Figure 4 The mechanical component and driver of the mirror alignment set: Fixed-point (1), Driver of actuators (3), Mechanical interface (triangle) (4), Actuators (5), Mounting sphere with telescope fixation (6).*

### 4.2 Actuators

The key element of the Mirror Alignment Fixing Set are actuators (see figure 5) that allow for changing the mirror facet orientation. There are two types of actuators, with 4 DoF and 5 DoF mounting joints. The design of the mounting joints allows transferring the actuator linear movement into mirror angular movement and also for compensation of the telescope structure thermal expansion.

*Development of an optical system for the SST-1M telescope of the Cherenkov Telescope Array observatory*

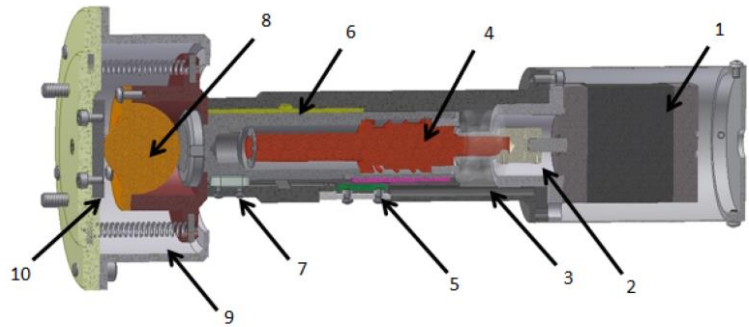

*Figure 5 Actuator design and components: Stepper motor (1), Oldham clutch (2), balls bearings (3), lead screw unit (4), magnetic encoder (5), linear bearing (6), sliding key (7), mounting joint (in case of 5DoF version - ball transfer unit (8), springs (90), plane (10).*

The operation principle of the actuator is based on the stepper motor which is connected with the screw lead unit through an Oldham clutch. The position of the actuator is controlled by a magnetic encoder which consists of a magnetic strip and integrated Hall sensor. The systems allows for precise linear motion with accuracy ~5μm in range of 20mm in CTA temperature operational range. The accuracy of the mirror spot positioning in focal plane is about 0.3 mm which fit well to expected mirror facet PSF (~4mm). The actuator can withstand axis loads up to 1200 N.

## 5. Telescope Optical Simulations

The components of optical system together with part of telescope structure was implemented in Zemax ® software. Later it was used for the analysis of optical point spread function (PSF) of the telescope. It was based on an analysis of enclosed energy to determine the telescope's nominal angular resolution. The analysis confirmed that the optical system meets the requirements – over 80% of energy was enclosed within <0.25°. The details related to the current SST-1M optical simulation can be found in [7].

The telescope mechanical structure i.e. camera casing and its supporting system allowed for determining the mirror effective area of the telescope dish. Analyses have shown that the mechanical supporting structure shades slightly more the mirrors facets and the effective area is equal to 80% instead of initially estimated 87%.

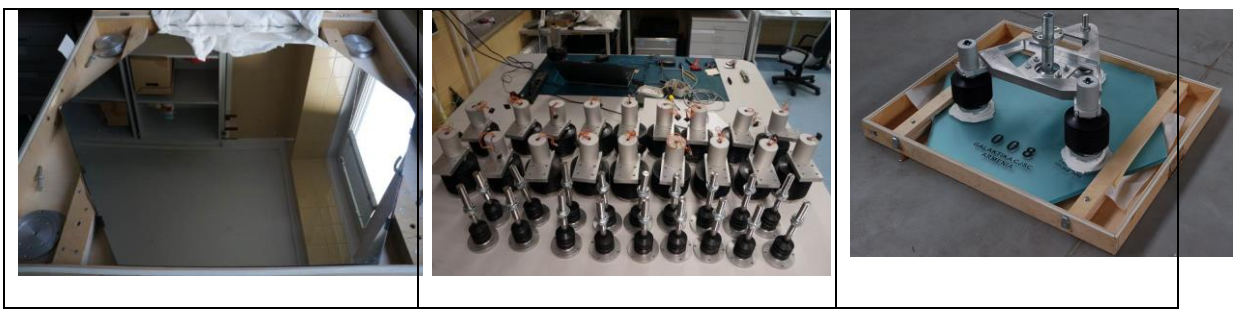

*Figure 6 Manufactured components of SST-1M Optical systems: glass mirror facet (left), actuators (middle), integrated mirror alignment fixing set (right)*

*Development of an optical system for the SST-1M telescope of the Cherenkov Telescope Array observatory*

## 6. Summary

Recently the component of SST-1M Optical System was manufactured (figure 6). The testing campaign is ongoing with aim to validate the designed parameters. In parallel the optical system was pre-integrated with telescope structure and it is presented on figure 7.

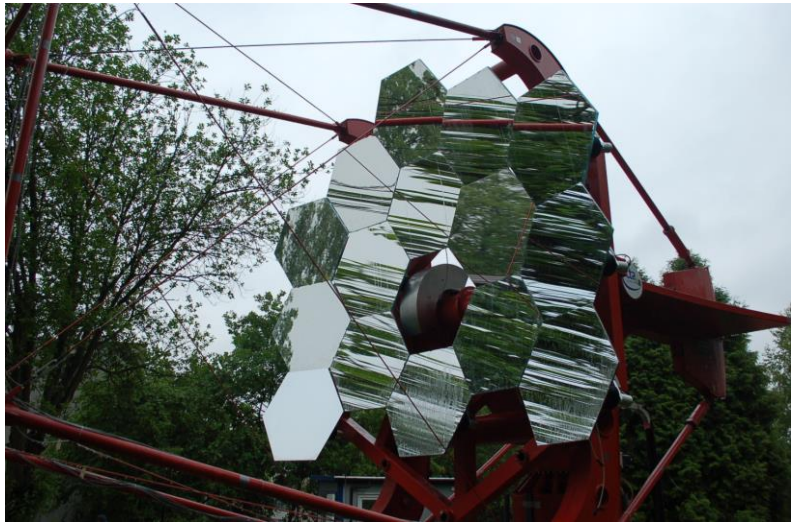

*Figure 7 Pre-integrated SST-1M optical system with telescope structure. Three remaining mirrors on the bottom of the telescope are plan to be integrated in autumn 2015.*

## Acknowledgments

We gratefully acknowledge support from the agencies and organizations listed under Funding Agencies at this website: http://www.cta-observatory.org/. In particular we are grateful for support from the Polish NCN grant DEC-2011/01/M/ST9/01891, the Polish MNiSW grant no. 498/1/FNiTP/FNiTP/2010, the University of Geneva, and the Swiss National Foundation.